\documentclass{article}

\usepackage[dvips]{graphicx}
\usepackage{amsmath,amssymb}

\begin{document}
\title{ Coherence length of cosmic background radiation enlarges 
the attenuation length of the ultra-high energy proton} 
\author{ K. Ishikawa and  Y. Tobita \\
\\
Department of Physics, Faculty of Science,\\
Hokkaido University Sapporo 060-0810, Japan}
\maketitle
\begin{abstract}
It is pointed out that an agreement of the one particle energy spectrum of 
the cosmic background radiation (CMBR) with  Plank distribution
of 2.725 [K] does not give a strong constraint on the coherence length of
CMBR if the mean free path of CMBR is very long. The coherence length in
this situation is estimated as a few times of $k_BT$.     Due to this  
finite coherence length,  the attenuation length of ultra-high energy cosmic 
rays (UHECR) is reduced in the  $\Delta $ resonance region,i.e.,  around 
$10^{20}$ [eV]. The small  attenuation length makes the
 suppression of the flux of cosmic rays in this energy region 
less prominent than the naive estimation.
\end{abstract}


\section{Introduction}

UHECR beyond $10^{20}$ [eV] is expected to be suppressed by its 
pion production collision  with CMBR by GZK bound. In this paper we
assume  that the
proton is UHECR and study its scattering with CMBR. Recently
observations of UHECR 
became possible by several experiments \cite{HaverahP,Fly's-eye,Lorentz,AGASA,Auger_pre} , and a possible signal of
UHECR around this energy region has been found, although experiments are 
controversial. A new mechanism which modifies the pion production
probability of UHECR is proposed in this letter.

CMBR has the temperature at around 2.7 [K] and is regarded as a wave
packet in the present work. Coherence is kept within the size of wave
packet, and we call it a coherence length.  
We show that the finite coherence length modifies the pion production
probability in the ${\Delta}$ resonance region.

When  the invariant mass of the initial
states composed of UHECR and a CMBR   
exceeds the pion production threshold,  the attenuation length
 of UHECR becomes short.  
GZK predicted the reduction of the flux beyond  this energy.
In the inelastic 
 cross section, the $\Delta$  resonance contribution is most important.  
The attenuation length  beyond this energy becomes the order of  20  [Mpc], which is less than
 the size of the universe. The UHECR can not propagate long distance in the 
 universe, then. So the flux of cosmic ray should be suppressed beyond 
this energy, known as GZK  bound.   For the estimation of GZK  bound, Lorentz
 invariance is assumed and the cross section between gamma and nucleon
 collisions in laboratory frame is used.  

Experiments in this  energy region are in progress and the  source of
UHECR will be identified in the future \cite{Auger}. Then  an
experimental determination of 
the attenuation length would become possible.

 We propose  a new mechanism of correction to  the GZK bound in this paper.
Most CMBR photons have been produced before the decoupling time in the 
early universe and may have finite coherence length. High energy charged
particle also has a finite coherence length due to the collinear mass 
singularity.
In the collisions of   CMBR with UHECR beyond  $10^{20}$ [eV] 
region, the finite coherence length makes the total photon-nucleon
energy to spread.
If this energy width is larger than or equivalent
to the width of the $\Delta$ resonance, the total amplitude 
in the $\Delta$ resonance region is suppressed. The attenuation 
length of UHECR is modified, then.  

 The effect of the finite coherence length is negligible in the ordinary
high energy scattering since the coherence length is much larger than the de
Broglie wave length. So this effect  has not been taken into account in the 
previous works on GZK bound. However, we show in the present work that
the finite coherence
length of CMBR gives a sizable effect to UHECR's attenuation length even 
though the CMBR spectrum agrees with
the Plank distribution.\cite{Stodolsky}


 \section{Wave packet and the coherence length}

In ordinary scattering experiments, a position where the beam particle 
is produced is known and the  wave packet size is determined from the
mean free path of the particle in matters  and its  size is
semi-microscopic with  much larger value  than de Broglie  length 
of high energy particle. Its effect is   negligible, then. Let us call this
length as the first coherence length. 
In a dilute system  we study, a position where the particle is produced is 
unknown and   the wave packet size, i.e., the coherence  length is
determined by the amplitude of many particle states in the
production process.

In a system where each particle has a large mean free path, particle
states preserve coherence for long time and coherence length due to mean
free path is negligible. In this region, particles are described by a many
body wave function and  coherence length due to many body effects
becomes important. Final state of the 
scattering matrix is a linear combination of the momentum states with the
weight of scattering amplitude, unless  a measurement
of the final state is made. Consequently a correlation for one particle 
state of the different momenta, which is
defined from the product between the scattering amplitude and its
complex conjugate, becomes finite. This   correlation
length due to many body wave functions becomes important in the dilute
system.

\begin{figure}
\centerline{\includegraphics[scale=.8]{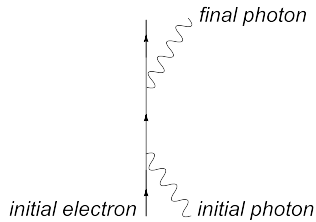}}
\caption{The diagram for Thomson scattering amplitude where the initial
 photon follows the Planck distribution and the initial electron follows the
 Fermi-Dirac distribution.}

\label{thomson}
 \end{figure}

We estimate this coherence  length of photon from the final states 
of Thomson scattering of Fig.~\ref{thomson}. Let us focus on the final
 scattering of CMBR.  
CMBR is composed of almost free photon with the Plank distribution as
far as single particle energy distribution is concerned. Thomson
scattering amplitude
is  independent from the scattering angle at the low energy. Also the
production region is not identified for CMBR.  Consequently 
the photon in the final states of Thomson scattering is a coherent linear 
combination of momentum states with different orientations. This  correlation
of photons with different  momenta is computed from these amplitudes 
where in the initial states the photon follows the  Plank 
distribution with arbitrary angle and the  electron follows the 
Fermi-Dirac distribution with arbitrary angle. We found that the
correlation of final photons   agrees   almost perfectly with that of
Gaussian wave packet\cite{Ishikawa-tobita-ptp}. Its width is a few times of $k_BT$. So even though 
photon follows the Plank distribution it has a correlation of Gaussian 
wave packet of the width of a few $k_BT$. We study its implications in the
collision of CMBR with UHECR.
 $\Delta$ resonance lives for short period and its amplitude has a finite
energy width. Hence if the time scale for UHECR to overlap with CMBR
is shorter than the life time of the $\Delta$ resonance, the amplitude 
is reduced  by  the finite coherence length.  Then  cross section due to 
the $\Delta$ resonance gets a sizable correction from  the finite
coherence length. 
 
Scattering amplitudes  for the wave
packets of  a finite spatial extension in relativistic field theory have 
been formulated in Ref.~\cite{Ishikawa-shimomura}. It is shown that the 
amplitudes are consistent with the probability 
interpretation,   despite non-orthogonality of the different states, and
that the conservation of the energy and momentum is satisfied within an
uncertainties given by the finite interaction area of the wave
packets, since
the states defined by wave packets have finite extensions. Furthermore,
the asymptotic condition is satisfied with an finite initial time $T_i$
and an finite final time $T_f$, and  
the scattering  probability has a position-dependence   in addition to  
a momentum-dependence. From these properties, the wave packet scattering 
amplitude is different from a simple linear combination of  
the plane wave amplitude.

Gaussian wave packets, i., e., coherent states,
of spherical symmetry  are  defined as
\begin{eqnarray}
\label{eq:coherentx}
& & \langle {\vec p}|{\vec P}_0,{\vec X}_0 \rangle = 
N_3{({\sigma_x^2})}^{\frac{3}{2}} e^{-i{{\vec p} \cdot {\vec X}_0} -{\sigma_x^2 \over 2}
({\vec P}_0-{\vec p})^2} 
~N_3^2=(\pi {\sigma}_x^2)^{-{3 \over 2}},{{\vec P}_0 \over \hbar}={\vec k}_0.
\end{eqnarray} 
Generalizations to the asymmetric wave packets and to non-minimum wave 
packets are straightforward.    
The set of functions for one value of $\sigma$ satisfy the
completeness condition. 

The time evolution of the free wave is determined by the free Hamiltonian,
and creation and annihilation operators which satisfy $[a({\vec p},t),a^{\dagger}(\vec{p}{\mspace{2mu}}^{\prime},t')]\delta(t-t')  =\delta(\vec{p}-\vec{p}{\mspace{2mu}^{\prime}})\delta(t-t')$
The operator $A({\vec P}_0,{\vec X}_0 ,T_0,t)$ 
that annihilates the state described by the wave packet is 
defined by  
\begin{eqnarray}
\label{eq:atoA}
A({\vec P}_0,{\vec X}_0,T_0,t )=\int d{\vec p}  \mspace{3mu}a({\vec p},t) \langle {\vec p
}|{\vec P}_0,{\vec X}_0,T_0\rangle  
\end{eqnarray}
and the creation operator is defined by its conjugate. The particle
states expressed by the wave packets follow classical trajectories and 
have finite spatial extensions. Consequently overlap region is finite
and the energy-momentum conservation is only approximate.
   
The wave packet spreads with time and the
spreading velocity in the transverse direction, ${v_T}$, and the
longitudinal direction, ${v_L}$,are  given by
\begin{eqnarray}
& &{ v}_T=\sqrt{{ 2\over \sigma_x^2}} { 1 \over E({\vec P}_0)},\ 
{ v}_L=\sqrt{{ 2\over \sigma_x^2}} { m^2 \over (E({\vec P}_0))^3}.
\end{eqnarray}

The ${v_T}$ depends on the energy and  the  ${v_L}$ depends on the
energy and the mass. A massive wave packet spreads
in both directions but a massless wave packet spreads only in the
transverse direction. After a macroscopic time, any wave packet of the
massive particle spreads to huge size. These wave may be treated as a
plane wave approximately. However  any wave packet of the massless
particle does not spread and its size is kept fixed in the longitudinal 
direction. Thus, the wave packet of massless particle remains for the
long period and its effect is important. 

\section{Resonance in the wave packet scattering}

In the scattering of high energy proton with CMBR, the square of the
center of mass energy, $S$, is defined by 
\begin{eqnarray}
 S  = (M_p^2 + 2E_pE_{\gamma} -
  2|\vec{P_p}|\vec{P_{\gamma}}|\cos{\theta}) \simeq (M_p^2 +
  2E_p\cdot E_{\gamma}(1 - \cos{\theta}))
\end{eqnarray}
where $(E_{\gamma}, {\vec P}_{\gamma}), (E_{p},{\vec P}_{p}),
\mbox{and} \ \theta $
are four momenta of the photon, of the proton and the collision angle. 
The mass and width of the $\Delta$, $M_{\Delta}$ and   $\Gamma $, 
are $ M_{\Delta} = 1232 [\mathrm{MeV}], \ \Gamma = 120 [\mathrm{MeV}]  $.

Breit-Wigner partial wave amplitude and the total cross section are,
\begin{eqnarray}
f_l({\theta})&=&{\sqrt{2l+1} \over p }{\Gamma/2 \over {\sqrt S}-
M_{\Delta}+i\Gamma/2 },~\sigma_l = \frac{4\pi(2l+1)}{p^2}
\frac{\left(\frac{\Gamma}{2}\right)^2} {(\sqrt{S} - M_{\Delta})^2  + 
\left(\frac{\Gamma}{2}\right)^2} .
\end{eqnarray}
The photon wave function  
of the momentum ${\vec p}_0$ which we obtained  is expressed by a
minimum wave packet,
\begin{eqnarray}
 |\psi_{\gamma}(\vec{p} - {\vec{p}}_0)| = N 
\exp\left[-\frac{(\vec{p} - {\vec{p}}_0)^2}{2\sigma^2}\right] \label{gamma-wave-function}.
\end{eqnarray}
 We have studied also non-minimum  case by multiplying a polynomial
$h_m({\vec p}) $ to the last term but our conclusion of the present work 
is the same. 
For an asymmetric wave packet, an asymmetric $\sigma$ is used. 
Actually a high energy charged particle is combined with coherent soft 
photons in order to cure infra-red
divergence\cite{infra-red_divergence,infra-red_divergence2} caused by
massless particle, photon. So a charged particle system is spread in the
momentum and energy.

\section{Cross section and attenuation length of UHECR}

The average cross section for the CMBR of finite coherence length is given by,
\begin{eqnarray}
& & \sigma_{CMBR} = \frac{\int d{ \nu} U({ \nu})\Bigg|\int d^3p \ \psi_{\gamma+p}(\vec{p}-{\vec p}_0)
  \frac{4\pi}{P_{CM}}
  \frac{\left(\frac{\Gamma}{2}\right)}{\sqrt{S} - M_{\Delta} +
  i\frac{\Gamma}{2}}\Bigg|^2}{\int d{ \nu} U({ \nu})},\label{CMBR-cross-section}\\
& & U({ \nu} ,  T) = \frac{8{\pi}h{{ \nu}}^3}{c^3}\frac{1}{e^{\frac{h{ \nu}}
{k_B  T}}-1},  |\vec{p}_0| =h { \nu} 
\end{eqnarray}
where $P_{CM}$ is momentum in the center of mass frame and parameters $h$,
$c$, $k_B$, $ T$, are Planck constant, speed of light, Boltzmann constant,
and temperature of the CMBR. $\psi_{\gamma+p}(\vec{p})$ is the wave
function of the photon-proton system. It  should be noted that the
integration on ${\vec p}$ is taken in  the amplitude because the
strict energy conservation does  not hold  for the  wave packet scattering. 



\begin{figure}
\centerline{\includegraphics[scale=.41,angle=270]{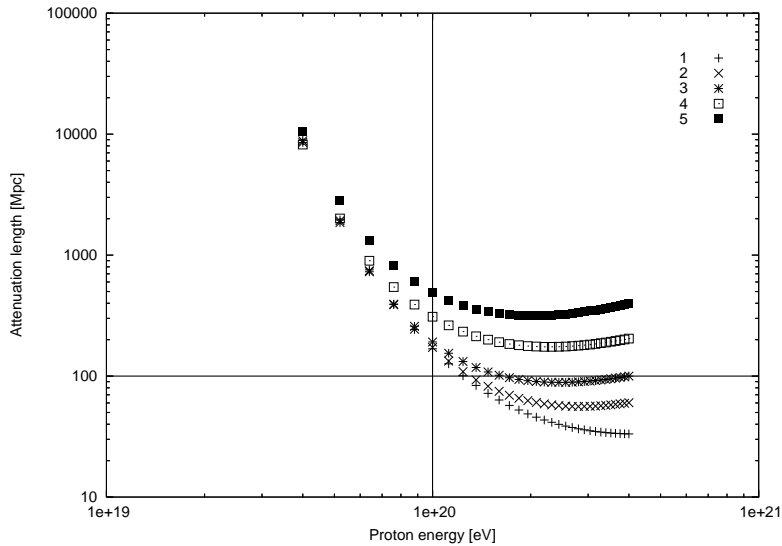}}
\caption{Proton energy dependence of attenuation length is given.\ {\bf 1} is for plane wave,\ {\bf 2} is for $\sigma =
  2.725[\text{K}](=T_{\text{CMBR}})$ ,\ {\bf 3} is for $\sigma =
  5.450[\text{K}]$,\ {\bf 4} is for $\sigma = 13.625[\text{K}]$ and {\bf5} is
  for $\sigma = 27.250[\text{K}]$.}
\label{fig:1}
 \end{figure}

Using the cross section, we calculate the average attenuation length of
UHECR in the parameter range $\sigma \leq 10 k_BT $, where the
temperature is regarded as that of the decoupling time until this
point. However to compare with the current observation, we use the current 
value of the temperature, $T= 2.725[\text{K}]$.  
The inelasticity of the UHECR is given by the ratio of energy loss of 
the UHECR, $E_{p,f}-E_{p,i}$, over the initial energy in the scattering  
$K_p={ E_{p,i}- E_{p,f} \over E_{p,i}}$.
The attenuation length is obtained by integrating the product of 
the above ratio, $K_p$, with  the cross section. The result is given in 
Fig.~\ref{fig:1}.
The attenuation length becomes longer by factor 10 if the coherent
energy spread is wide. Since the $\sigma$ includes both effects of CMBR
and UHECR,  larger values of the $\sigma$
is included  here.  

\section{Lorentz invariance} 

By a Lorentz transformation of the frame, one value of the momentum is 
transformed to another value of the momentum \cite{Lorentz-breaking1,
Lorentz-breaking2,Lorentz-breaking3}. Amplitude for the plane
wave is covariant and the cross section is invariant under the Lorentz 
transformation. So the cross section for the plane wave for CMBR is
computed easily from the experimental value of the photon-nucleon
reactions  in  the laboratory frame.  The amplitude for the wave packets, however, should be treated carefully, since the wave packet is a
linear combination of the momentum states.

We calculated the total cross section for the wave packet explicitly and 
find the sizable difference compared to the plane wave.  The finite size 
effect of the photon is negligible in the laboratory frame of  the pion 
production process in photon-nucleon reactions but is 
not negligible in the present situation. In the latter system, a small
coherent energy spread of CMBR leads the center of
mass energy to spread finite amount. A product of the small coherent
energy spread with an extremely large energy of UHECR becomes finite.
The variance of center of mass energy
 $S$ of the laboratory frame, ${\Delta S}_l$, and of the CMBR frame,
 ${\Delta S}_{CMBR}$, are given by
\begin{eqnarray}
{\Delta S}_l&=&2 m_p {\Delta E}_l,\\
 {\Delta S}_{CMBR}&=&2 p {\Delta E}_{CMBR} \delta (1-a),
\end{eqnarray}
where ``a'' is of order $1$.  If the energy variances of the photon are the same in both systems, the ratio of two values of the center of mass
energy is given by $p \over m_p$. This ratio is  of order $10^{11}$ for UHECR.
So in this case $\Delta S_{CMBR}$ is much larger than $\Delta S_{l}$
and  the finite coherent length effect in the CMBR frame of UHECR is much 
larger than that of  the laboratory frame. This shows the
reason why a naive  Lorentz invariance does  not hold  for the wave
packet scattering.

\section{Summary}

We found that the attenuation length of UHECR in the $10^{20} $ [eV] region
varies depending on the coherence length of CMBR. It is clear 
from the Fig.~\ref{fig:1} that  the attenuation length for UHECR becomes longer 
if the CMBR photon has a finite coherence length.  The effect becomes
important in
the pion production threshold energy  region, where a new data from Auger
collaboration concluded that the UHECR in this energy region comes
sources within 75 [Mpc] \cite{Auger2}\cite{Auger3}. Our calculation at the coherence length of $3.5
k_BT$ suggests that the value becomes 150 [Mpc] instead of 75 [Mpc].
When the precise value of the flux of UHECR in
wider energy region will be known, better informations will be
obtained.

Our study  shows the importance of the finite 
coherence length of CMBR in analyzing  the attenuation length of
UHECR. 

\section*{Acknowledgements}

This work was partially supported by the special Grant-in-Aid
for Promotion of Education and Science in Hokkaido University
provided by the Ministry of Education, Science, Sports and Culture,
 a Grant-in-Aid for Scientific Research (Grant No. 19540253), and a 
Grant-in-Aid for Scientific Research on Priority Area ( Progress in Elementary
Particle Physics of the 21st Century through Discoveries  of Higgs Boson and
Supersymmetry, Grant No. 16081201) provided by 
the Ministry of Education, Science, Sports and Culture, Japan.
\\
\\
\\

\end{document}